# The Doppler shift of resonant fluorescence spectrum for a two-level $^{85}Rb$ atom via multi-photon Compton Scattering

C$_{HAO}$ Y$_{ING}$ Zhao,[1,2,*] W$_{EI}$ Fan,[3] W$_{EI}$H$_{AN}$ Tan[4]

[1]*College of Science, Hangzhou Dianzi University, Zhejiang 310018, China*

[2]*State Key Laboratory of Quantum Optics and Quantum Optics Devices, Institute of Opto-Electronics, Shanxi University, Taiyuan 030006, China*

[3]*Shanghai Institute of Optics and Fine Mechanics, Chinese Academy of Sciences, Shanghai, 201800, China*

[4]*Department of Physics, Shanghai University, Shanghai 200444, China*

*Corresponding author: zchy49@163.com

Usually, it's difficult for us to observe the Compton Scattering in an atom. One way to overcome this difficult is using multi-photon collide with an atom, which will come into being multi-photon Compton Scattering (MCS) phenomenon. Thus, we can investigate the MCS process in visible light region. During the MCS process, the cluster atoms moving as a whole, namely atomic Dicke states, the multi-photon interacting with cluster atoms. We can observe a significant Doppler shift of resonant fluorescence spectrum(RFS)in a room-temperature two-level $^{85}Rb$ atomic system. In this paper, we present a detail analysis of the physics mechanism of the Doppler shift and propose a method to measure the component of the Dicke states (the atomic polymers with different masses)by using the Doppler shift of the RFS.

## 1. INTRODUCTION

The Compton Scattering phenomenon was firstly observed when a $\gamma$ ray collides with an electron[1].The Compton wavelength of the electron is $\lambda_c = h/m_e c$, where $m_e$ is the electron rest mass, $h$ is the Planck constant, $c$ is the speed of light. Substituting $\gamma$ ray by photon, we may investigate a photon collides with an atom, where $m$ is the atomic mass. Due to $h/mc \ll h/m_e c$, it is very difficult to observe the single photon Compton Scattering process. The only way to overcome this difficult is using multi-photon collides with an atom. The Doppler shift associated with the collision of atoms with multi-photon, the magnitude is dependent on the direction of observation and the mass of atoms. The physics mechanism and magnitude of the Doppler shift is quite different from that of the Doppler broadening. The Doppler broadening associated with the thermal motion of atoms, the magnitude is independent on the direction of observation due to isotropic.

Since the invention of the laser in the early 1960s, the laser beam is a kind of electromagnetic wave which has high density of photons that may be in coherent states, the photon degeneracy up to $n = 5 \times 10^7$ [2-4]. Thus, we can observe the MCS process in visible light range. Considering many photons collision simultaneously with cluster atoms, namely the Dicke states[5-6], which moving as a whole, the MCS process produce energy $nh\nu$ and momentum $nh\bar{k}$. The MCS can be represented by the Feynman diagram where the electron is treated as an undressed one, the result can be applied only to a low intensity laser. In order to investigate the effect of the coherency of the photon on Compton Scattering, the formula based on coherent-state formalism is required.

When a two-level atom interacting with a single-photon wave packet, the atom absorb one photon transits from the ground state to the excited state, and then return to the ground state again, radiates resonance fluorescence spectrum (RFS). The power spectrum of the light scattered by a two-level atom in the presence of a coherent continuous-pulse-train driving field is analyzed. Separate expressions for the coherent and incoherent components of the spectra are obtained[8-9]. The coherent part of the spectrum consists of a very narrow peak, without energy and momentum exchange during the collision of atoms and photon, The incoherent spectrum has the energy-momentum exchange process. That is, the spectrum consists of a series of triplets in RFS[10-11]. We consider a dense two-level $^{85}Rb$ atomic gas, supposed in equilibrium at room temperature $T$. There are two kinds of phenomenon happened in RFS: One is the random thermal motion of the radiating atoms induced Doppler broadening(isotropic), the magnitude is independent on the direction of observation.

$(\Delta v/v)_{Db} = \sqrt{2kT/m_{Rb}}/c \approx 8.5 \times 10^{-7}$ is very small, may be neglected. Another is the collision of atoms with photons produce Doppler shift, the magnitude is depend on the direction of observation and the atomic mass. Guo firstly investigated the Doppler shift and observed the recoil-induced of resonances[12]. Stark firstly detect the Doppler shift for light emitted by the moving hydrogen atoms[13]. Recently, Utsunomiya produced the quasi-monochromatic $\gamma$ ray by the collision of photons with relativistic electrons[14]. Although the RFS has a quiet complicate structure[15-16], there are a significant Doppler shift $(\bar{v} - v)/v \approx 1.18 \times 10^{-3}$ can be observed. In this paper, we will investigate the Doppler shift and the Dicke states via the MCS process. The organization as follows. Firstly, we discuss MCS process, secondly, we investigate the RFS and DRFS (RFS with Doppler shift) of two-level atomic system. Thirdly, we investigate the RFS and DRFS of atom Dicke state. Finally, Conclusions are given.

## 2.THE MULTI-PHOTON COMPTON SCATTERING PROCESS

The incident photon moves along the $x$ axis, the photon number states $|n\rangle$, the eigen-value $n$ unchanged in the during of Scattering. We make the assumption that the atom with mass $m$ stay in the static state, the initial velocity of atom is zero, the energy and momentum exchange satisfied the conservation equations of the participated particles[17]. Fig.1 shows the single-photon Compton Scattering process. Suppose $v$ is the frequency of the photon before the collision, $\bar{v}$ is the frequency of the scattered photon after the collision, the angle $\alpha$ between the direction of motion of the photon and the direction of motion of the scattered photon, the angle $\beta$ between the direction of motion of the photon before and after the collision, that is, the scattering angle.

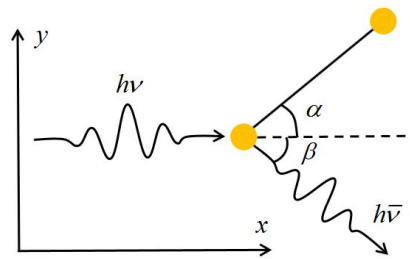

Fig. 1. The single-photon Compton Scattering process.

Einstein think that the quantum '$h\nu$' always carries a momentum '$h\nu/c$', the emission of a quantum '$h\nu$' by an atom produces a "jump" in its velocity and that this jump is responsible for the Doppler shift. A detailed account is presented of the Doppler effect as a photon phenomenon[18-19]. Schrödinger think that the emission of a light quantum by a moving atom satisfied the conservation laws of energy and momentum [20]. $n$ photon collisions simultaneously with atoms, which is moving with the same velocity as the system as a whole along the $x$ axis, substituting '$h\nu$' by '$nh\nu$'.

When a laser field is applied to an atomic system, the nonlinear Compton effect will appear, in which several photons are absorbed in a single point, not only a single high-energy photon is emitted, but also an atom possible scatters twice or more as it traverses the laser focus. Assuming that atom and photons produce nonlinear MCS, the frequency of Compton Scattering photons increasing with the number of photons $n$. The single photon energy of laser is far less than the static energy of atom, the laser field can be regarded as the classical wave field, we can deal with the effect of atom and light field in the range of relativistic electrodynamics. At beginning, the photon is moving, it is difficult to determine the behavior of the Scattering photon. So we assume the atom with mass $m$ is to be stationary before collision. The momentum of the atom and $n$-photon before Scattering is $mc^2$ and $nh\nu$, respectively. After Scattering, the momentum of the atom and the emitted high-frequency photon are $mc^2/\sqrt{1-u^2/c^2}$ and $nh\bar{\nu}$, respectively. The individual probabilities of emissions are not independent of each other, because the initial atom momentum at each emission is different. In fact, the emission of each photon modifies the atom state and consequently the next emissions. If the polarization of the photon is not considered, according to the conservation of the momentum, when the photon is scattered by the atom, there are

$$mc^2 + nh\nu = \frac{mc^2}{\sqrt{1-u^2/c^2}} + nh\bar{\nu},$$

$$\frac{nh\nu}{c} = \frac{nh\bar{\nu}}{c}\cos[\beta] + \frac{mu}{\sqrt{1-u^2/c^2}}\cos[\alpha],$$

$$0 = -\frac{nh\bar{\nu}}{c}\sin[\beta] + \frac{mu}{\sqrt{1-u^2/c^2}}\sin[\alpha], \quad (1)$$

Before the collision, the photons moving in parallel with the $x$ axis. After the collision, the photons moving in direction making an angle $\beta$ with the $x$ axis, the atom with a velocity $u$ in direction making an angle $\alpha$ with the $x$ axis.

$$mn(\nu - \bar{\nu}) - \frac{h}{c^2}(1-\cos[\beta])n^2\nu\bar{\nu} = 0,$$

$$\bar{\lambda} - \lambda = \frac{c}{\bar{\nu}} - \frac{c}{\nu} = 2n\lambda_c \sin^2[\frac{\beta}{2}], \quad (2)$$

Eq.(2) shows that the MCS wave length $\lambda_{nc} = nh/mc = n\lambda_c$, the photon degeneracy is increased by $n$ time compared with that of the single-photon Compton wave length $\lambda_c$. There are two special solutions: One solution $\beta = \pi$ represents the mirror image reflection, the wave length difference $\bar{\lambda} - \lambda = 2n\lambda_c$ attains a maximum after colliding with atom. The other trivial solution $\beta = 0$ represents the photons being absorbed and initiating the resonant florescence. In our model, a vast majority of photons used for the mirror image reflection, only a small part of photons initiating the resonate florescence.

When $\beta = \pi$, $\alpha = 0$, we have the following relations

$$\cos[\alpha] = 1, \quad \cos[\beta] = -1,$$

$$\bar{\lambda} = \lambda + 2n\lambda_c, \quad \lambda_c = \frac{h}{mc}, \quad \nu_c = \frac{c}{\lambda_c}, \quad \nu = \frac{c}{\lambda}, \quad \frac{\bar{\nu}}{\nu} = \frac{\lambda}{\lambda + 2n\lambda_c} = \frac{1}{1 + 2n\nu/\nu_c},$$

$$\frac{nh\nu}{c} + \frac{nh\bar{\nu}}{c} = \frac{mu_n}{\sqrt{1-u_n^2/c^2}}, \quad \frac{u_n}{c}\frac{1}{\sqrt{1-u_n^2/c^2}} = \frac{n\hbar(\nu+\bar{\nu})}{mc^2} = A, \quad (3)$$

The Doppler shift takes the form[21]

$$\frac{\bar{\nu} - \nu}{\nu} = \frac{u_n/c}{1 - u_n/c} \approx \frac{A}{1-A} \approx A. \quad (4)$$

In general, the multi-photon interaction with cluster atoms(namely Dicke states) cooperatively interaction with a single mode radiation field. The number of Dicke states $r$, here $r = 1, 2, 3, ...$ represents monomer (single atom $r = 1$), dimer (two atoms $r = 2$),..., polymer (multi-atoms $r = n$), respectively. The mass of proton $m_p = 1.672 \times 10^{-24} g$, for a monomer, $m_{Rb} = 76 \times m_p = 1.27 \times 10^{-22} g$; for a dimer, $2m_{Rb} = 2.54 \times 10^{-22} g$. In general, $r$ atoms in the polymer state, moving as a whole, $rm_{Rb} = r \times 76 \times m_p$, multi-photon interact with $r$ atoms.

For a monomer,

$$\lambda_c = \frac{h}{mc} = 1.739 \times 10^{-15} cm, \quad \frac{v}{v_c} = v\frac{\lambda_c}{c} = v\frac{1.739 \times 10^{-15} cm}{3 \times 10^{10} cm/s} = 0.58 \times 10^{-25} v\, s, \quad (5)$$

The frequency of visible light $v = 10^{15} s^{-1}$, we have $v/v_c = 0.58 \times 10^{-10}$. If the photon degeneracy $n = 10^7$ [10-11], we have $nv/v_c = 5.8 \times 10^{-4}$.

Substituting Eq.(4) into Eq.(5), we have the MCS Doppler shift

$$\frac{\bar{v} - v}{v} = \frac{nv}{v_c}(1 + \frac{1}{1 + 2nv/v_c}) \approx 1.16 \times 10^{-3}, \quad (6)$$

After Scattering, the atom moves towards observer, the resonance fluorescence frequency $v$ radiated by the atom, and the Doppler shift turns to $v u/c/(1 - u/c)$, the observed frequency $v_{ob}$ becomes

$$v_{ob} = v(1 + \frac{\bar{v} - v}{v}) = v(1 + \frac{nv}{v_c}(1 + \frac{1}{1 + 2nv/v_c})), \quad (7)$$

The Doppler shift attains to $v_{ob} - v \approx 5.8 \times 10^{-4} v$, it is significant.

We note that the MCS Doppler shift in Eqs.(6)-(7) contain two terms in the bracket, the first term $nv/v_c$ defines by incident photon momentum, the second term $nv/v_c(1 + 2nv/v_c)$ represents reflected photons. Considering the atom's transition from the ground state to the exited state, one of the reflected photons is absorbed, the Eqs.(6)-(7) should be modified as

$$\frac{\bar{\nu}-\nu}{\nu} = \frac{n\nu}{\nu_c}(1+\frac{(n-1)}{n}\frac{1}{1+2(n-1)\nu/\nu_c}), \tag{6'}$$

$$\nu_{ob} = \nu(1+\frac{n\nu}{\nu_c}+\frac{(n-1)\nu}{\nu_c}\frac{1}{1+2(n-1)\nu/\nu_c}), \tag{7'}$$

When $\Delta t > 1/\Delta \nu$, according to Eq.(7) and Eq.(7)', the calculation result show that the curves essentially indistinguishable.

When the transition happens, the atom will stay in the excited state until $\Delta t > 1/\Delta \nu$, namely out-of-photon-state, where $\Delta \nu$ is spontaneous emission bandwidth of the atom, Here, we consider a two-level system and the MCS is a coherent Scattering. When the atom transits to the ground state and absorbs the second photon, and then transits to the excited state. But the longitude mode $\Delta t \Delta \nu \leq 1$ restricts transition to the next mode, which violation of the coherent Scattering. therefore two-photon absorption never happen in the same photon state ( $\Delta t \Delta \nu \leq 1$ ).One way to absorb more photons is to change two-level to multi-level system, thus more photons are absorbed, for example, $k \geq 2$. In this case, the second term in Eqs.(6)'-(7)' may be replaced by

$$\frac{(n-k)\nu}{\nu_c}\frac{1}{1+2(n-k)\nu/\nu_c}.$$

When $n \gg k$, the influence on calculation results is still small.
Another way of absorption more photons is substituting the coherent MCS longitude mode $\Delta t \Delta \nu \leq 1$ restriction by the incoherent MCS. Before and after collision, not in the same photon state, the conservation of energy and momentum does not hold, the photon degeneration is different. It is very difficult to get an analytical solution for the incoherent MCS. The strong nonlinear MCS possible occurs in a high atoms and ions density laser plasma, collision between atoms and ions play an important role also. The physics mechanism is very complex beyond the coherent MCS considered in the present paper. For the coherent MCS, the atom density is very low, therefore the collision between atoms can be neglected.
As shown in Fig.2, the RFS frequency $\nu'_{ab}$ observed perpendicular to the direction of the atoms' motion, and the Doppler shift DRFS frequency $v_{ob}$ observed in the direction of the atoms' motion.

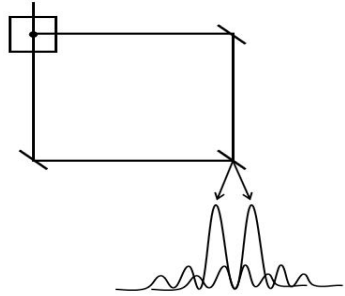

**Fig. 2.** The scheme of multi-photon Compton Scattering.

## 3.THE RFS FOR TWO-LEVEL ATOMIC SYSTEM DRIVEN BY THE COHERENT LIGHT

As we all known, the RFS of a single two-level atom contains three peaks (Mollow triplet)[8-9], and the RFS experiments have been realized by Wu and Grove[10-11], and Guo[12]. For the RFS of multi-atoms system situation, Senitzky[22] predicted that there are many side peaks, such as $\omega = \omega_0 \pm n\Omega$, $n = 1, 2, 3,...$, where $\Omega$ is the Rabi frequency, besides the resonance peak. However, Argwal [23-24] studied the RFS of two atoms system by solving the density matrix dynamical equations. When driven light intensity is not very high, the RFS possess $\omega = \omega_0$, $\omega_0 \pm \Omega$ three peaks, while when driven light intensity is very high, the $\omega_0 \pm 2\Omega$ peaks appear. Tan first noticed that the emission of one atom modified the electron states of the neighborhood atom's emission and solved theoretically the two atoms density matrix dynamical equations[16]. The RFS gradually transits from three peaks(single atom RFS) to five peaks when the emission modified neighborhood atom coupling constant $\alpha_s = 0, -1, -2, -3$, respectively. But this model is not suitable for Dicke states which is defined that the atoms only couple to a common electromagnetic field $E$, there is no direct interaction between the atoms. The emission modified neighborhood mutual coupling belongs to the direct interaction between the atoms, violation of the definition of the Dicke states should be abandoned. Moreover, even if we adopt a very complicated system in our calculation model, the corresponding reference spectrum (without Doppler shift) and moving spectrum(with Doppler shift) will change, but the relative shifts between two spectra unchanged. Now, we adopt Mollow's model as our calculation model.

Let us first consider the case in which the incident field frequency $\nu$ is exactly on resonance with the atomic transition frequency $\omega_{10}$, namely $\nu = \omega_{10}$, the coherent spectrum $g(\nu)_{coh} \propto \delta(\nu - \omega_{10})$ (where $\delta$ is the Dirac-function), and the incoherent spectrum $g(\nu)_{incoh}$ namely the RFS[8-9]

$$g(\nu)_{incoh} = \frac{\kappa \bar{n}_\infty \Omega^2 ((\nu - \omega_{10})^2 + (\Omega^2/2 + \kappa^2))}{((\nu - \omega_{10})^2 + s_0^2)((\nu - \omega_{10})^2 + s_1^2)((\nu - \omega_{10})^2 + s_2^2)},$$

and the two roots are given by

$$s_0 = -\kappa/2, \quad s_1 = -3\kappa/4 - \sqrt{\kappa^2/16 - \Omega^2}, \quad s_2 = -3\kappa/4 + \sqrt{\kappa^2/16 - \Omega^2}, \quad (8)$$

where the average value of photons $\bar{n}_\infty = (\Omega^2/4)/(\Omega^2/2 + \kappa^2/4)$ approaches equilibrium value[1], $\kappa$ is the spontaneous emission coefficient, the definition of notations $\bar{n}_\infty$, $\kappa$ refer to Ref.[9]. $\Omega = 2\mu E_0/\hbar$ is the Rabi frequency of pump field, $E_0$ is two-level atom energy, $\mu$ is the electric dipole moment. As long as we know the spectral distribution represented by the atomic operator, we can determine the statistical distribution law of $n$-order correlation function of the fluorescence field, furthermore we can determine the state function of the resonance fluorescence field. As we all known, a laser can be well described by a coherent state $|\alpha\rangle$. When the phase of a coherent state is randomized, it is equivalent to a mixed state of Fock states[25]

$$|\alpha\rangle = \sum_{n=0}^{\infty} e^{-\frac{\alpha^2}{2}} \frac{\alpha^n}{\sqrt{n!}} |n\rangle = \sum_{n=0}^{\infty} \sqrt{p(n)} |n\rangle, \quad (9)$$

Where $\alpha$ is the coherence coefficient, the probability of $n$-atom obeys the binomial probability distribution $p(n)$. The normalization condition reads $\sum_{n=0}^{\infty} p(n) = 1$.

The number states are the eigen-states of the number operator $\hat{a}^\dagger \hat{a}$: $\hat{a}^\dagger \hat{a}|n\rangle = n|n\rangle$, $n = 1, 2, 3...$, where $\hat{a}$ and $\hat{a}^\dagger$ are annihilation and creation operators satisfying the commutation relation. The coherent states whose photon number follows Poisson distribution[25] with a mean of $|\alpha|^2$ [26].

$$\langle n \rangle = \langle \alpha | \hat{a}^\dagger \hat{a} | \alpha \rangle = e^{-\alpha^2} \sum_{n=0}^{\infty} \frac{\alpha^{2n}}{(n-1)!} = \alpha^2 \sum_{n=0}^{\infty} p(n) = \alpha^2, \qquad (10)$$

According to Eqs. (6)-(9), the Doppler shift of the RFS can be rewritten as

$$I(\nu) = g(\nu_{ob}) = \sum_{n=0}^{\infty} p(n) g_{incoh}((1-\frac{u_n}{c})\nu) = \sum_{n=0}^{\infty} p(n) g_{incoh}((1-A_n)\nu). \qquad (11)$$

In the following numerical simulation, we take the following parameters: the transition frequency $\omega_{10} = 386 THz$, the emission frequency $\nu = 385.5\text{-}386.5 THz$, the spontaneous emission coefficient $\kappa = 0.05 THz$, the Rabi frequency $\Omega = 3\kappa$ [27-28].

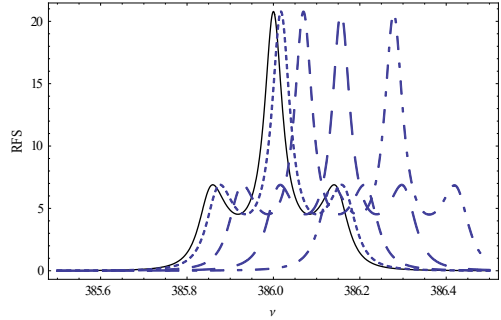

**Fig. 3.** Given the number of cluster atoms $r = 1$ (monomer), the RFS of $^{85}Rb$ atom (without the Doppler shift) with the coherence coefficient $\alpha = 0$ (the solid line), the DRFS of $^{85}Rb$ atom (with the Doppler shift) with the coherence coefficient $\alpha = 10^3$ (the narrow dotted line), $\alpha = 2 \times 10^3$ (the medium dotted line), $\alpha = 3 \times 10^3$ (the wide dotted line), $\alpha = 4 \times 10^3$ (the dash dotted line), respectively.

In Fig.3, we investigate the RFS and DRFS of $^{85}Rb$ atom for monomer via MCS. We find that the curve moves to right (blue shift) with the increasing of the different coherence coefficient $\alpha$. According to Eq.(9), the degeneration of the coherent state is $\langle n \rangle = \alpha^2$.

## 4. THE PHYSICAL DIAGNOSIS OF THE DICKE STATE

The single-photon Compton wavelength $\lambda_c$ is inversely proportional to the atomic mass $m$. The mass of cluster atoms should be $rm$, the cluster atoms cooperatively interaction with a single mode radiation field. For the Dicke states, the Compton wavelength $\lambda_c = h/mc$, $\nu_c = c/\lambda_c$ in Eq. (3) should be modified to

$$\lambda_{rc} = h/rmc = \lambda_c/r, \quad \nu_{rc} = c/\lambda_{rc} = r\nu_c,$$

$$\frac{u_{rn}}{c} \frac{1}{\sqrt{1-u_{rn}^2/c^2}} = n\frac{\nu}{r\nu_c}(1+\frac{1}{1+2n\nu/r\nu_c}) = A_{rn},$$

Accordingly, Eq. (11) turns to

$$I(\nu) = g(\nu_{rob}) = \sum_{n=0}^{\infty} p(n)g_{incoh}(\nu(1-A_{rn})), \tag{12}$$

Fig. 4 shows the RFS and the DRFS of $^{85}Rb$ atom via MCS with the same coherence coefficient $\alpha = 4\times 10^3$ and the different number of cluster atoms $r = 1-3$.

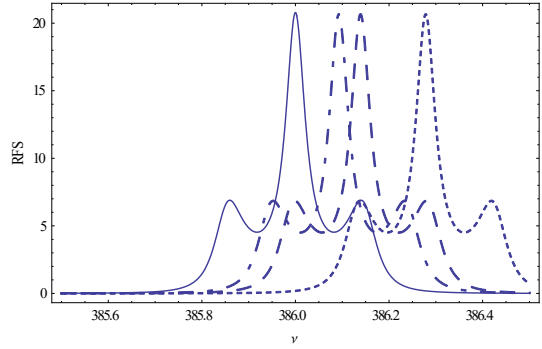

**Fig. 4.** The RFS of $^{85}Rb$ atom with the coherence coefficient $\alpha = 0$ (without the Doppler shift) and the number of cluster atoms $r = 1$ (monomer, the solid line). Given the coherence coefficient $\alpha = 4\times 10^3$, the DRFS of $^{85}Rb$ atom (with the Doppler shift) with the number of cluster atoms $r = 1$ (monomer, the narrow dotted line), $r = 2$ (dimer, the wide dotted line), $r = 3$ (trimer, the dashed dotted line), respectively.

## 5. CONCLUSION AND DISCUSSION

In this work, our attention is concentrated to determine the Compton process in Dicke states and analyze the generated Doppler shift in the RFS. The Dicke states is a coherent state of multiple particles, which is fundamentally responsible for various intriguing collective behaviors of many-body systems. The model we have adopted as the basis of our analysis consists of a cluster of two-level atoms driven by a laser field. The numerical solutions have been used to show the Doppler shift induced by MCS in the proposed model. In order to use MCS for the diagnosis of the Dicke states, the numerical solutions are depicted in Figs.3-4. There are some key points need to be mentioned:

(i). For a two-level ($|0\rangle$ and $|1\rangle$) atomic system, we choose the resonant transition line $\omega_{10} = 386 THz$ in Fig.3[27], the RFS is represented as $g_{incoh}(\nu)$ in Eq.(8), and the corresponding DRFS is expressed by $g_{incoh}(\nu_{ob})$ in Eq.(12). While for a three-level ($|0\rangle$, $|1\rangle$ and $|2\rangle$) atomic system, the influence of neighbor energy level $|2\rangle$ must be considered, the corresponding resonant transition line turns to $\omega_{20} = 388 THz$ [27-28]. Eq. (8) should be modified by $\bar{g}_{incoh}(\nu)$, and the corresponding Eq.(12) turns to $\bar{g}_{incoh}(\nu_{ob}) = \sum_{n=0}^{\infty} p(n) \bar{g}_{incoh}(\nu(1 - A_n))$.

(ii). In Figs.3-4, the RFS(the black line) is the reference spectrum. Besides the difference between the Doppler shift, furthermore we may investigate the resonant spectra difference between monomer and polymers[15-16]. In experiment, in the case of without of influence of Doppler shift, the observed resonant spectrum actually is the superposition of different composition ratio of the resonant spectra from monomer to polymers. If we consider the Doppler shift, the resonant spectrum separately from each other, we called it the Dicke states[5,6], the resonant spectra of the Dicke states have been studied from monomer, dimer to polymers[13,15]. Thus, the observation of a Doppler-shift resonant spectra give us a proper diagnosis that fit for the Dicke states, the constitution of the Dicke states and the associated resonant spectrum.

**Funding.** This work was supported in part by the National Natural Science Foundation of China (grant numbers11504074) and the State Key Laboratory of Quantum Optics and Quantum Optics Devices, Shanxi University, Shanxi, China (grant numbers KF201801).

**Disclosures:** The authors declare no competing interests.

**REFERENCES**

1. C. H. Yu, R. Qi, W. T. Wang, J. S. Liu, W. T. Li, C. Wang, Z. J. Zhang, J. Q. Liu, Z. Y. Qin, M. Fang, K. Feng, Y. Wu, Y. Tian, Y. Xu, F. X.Wu, Y. X. Leng, X. F. Weng, J. H.Wang, F. L. Wei, Y. C. Yi, Z. H. Song, R. X. Li, Z. Z. Xu, "Ultrahigh brilliance quasi-monochromatic MeV $\gamma$-rays based on self-synchronized all optical compton scattering," Scientific Reports **6**, 29518(2016).

2. T. H. Maiman, "Stimulated optical radiation in Ruby," Nature **187**, 493-494(1960).

3. L. Mandel, "Photon Degeneracy in Light from Optical Maser and Other Sources," J. Opt. Soc. Am. **51**, 797-798(1961).

4. R. J. Collins, D. F. Nelson, A. L. Schawlow, W. L. Bond, C. G. B. Garrett, W. Kaiser, "Coherence, Narrowing, Directionality, and Relaxation Oscillations in the Light Emission from Ruby," Phys. Rev. Lett. **5**, 303-305(1960).

5. R. H. Dicke, "The effect of collisions upon the Doppler width of spectral lines," Phys. Rev. **89**, 472-473(1953).

6. R. H. Dicke, "Coherence in Spontaneous Radiation Processes," Phys. Rev. **93**, 99-110(1954).

7. B. Zhang, Z. M. Zhang, Z. G. Deng, W. Hong, J. Teng, S. K. He, W. M. Zhou, Y. Q. Gu, "Multi-photon effects in single nonlinear Compton Scattering and single nonlinear Breit-Wheeler process in ultra intense fields," arXiv.1707.04841v2(2017).

8. B. R. Mollow, "Power spectrum of light scattered by two-level system," Phys. Rev. **188**, 1969-1975 (1969).

9. B. R. Mollow, "Theory of intensity dependent resonance light scattering and resonance fluorescence," Prog. Opt. **19**, 1-43(1981).

10. F. Y. Wu, R. E. Grove, S. Ezekiel, "Investigation of the Spectrum of Resonance Fluorescence Induced by a Monochromatic Field," Phys. Rev. Lett. **35**, 1426-1428(1975).

11. R. E. Grove, F. Y. Wu, S. Ezekiel, "Measurement of the spectrum of resonance fluorescence from a two-level atom in a intense monochromatic field," Phys. Rev. A **15**, 227-233(1977).

12. J. Guo, P. R. Berman, B. Dubetsky, G. Grynberg, "Recoil-induced resonances in nonlinear spectroscopy," Phys. Rev. A **46**, 1426-1428(1992).

13. J. Stark, "On the Radiation of Canal Rays in Hydrogen," Astr. J. **159**, 25(1907).

14. H. Utsunomiya, T. Watanabe, T. Ari-izumi, D. Takenaka, T. Araki, K. Tsuji, I.

Gheorghe, D. M. Filipescu, S. Belyshev, K. Stopani, D. Symochko, H. W. Wang, G. T. Fan, T. Renstrom, G. M. Tveten, Y. W. Lui, K. Sugita and S. Miyamoto, "Photon-flux determination by the Poisson-fitting technique with quenching corrections," Nucl. Instrum. Methods Phys. Res. Sect. A **896** 103(2018).

15. G. S. Agarwal, L. M. Narducci, E. Apostolidis, "Effects of dispersion forces in optical resonance phenomena," Opt. Commun. **36**, 285-290(1981).

16. W. H. Tan, M. Gu, "Resonance fluorescence in a many-atom system," Phys. Rev. A **34**, 4070-4078(1986).

17. P. G. Bergmann, *Introduction to the theory of relativity,* (New Jersey, Prentice-Hall, 1958).

18. G. Giuliani, "Experiment and theory: the case of the Doppler effect for photons," Eur. J. Phys. **34**, 1035-1047 (2013).

19. D. V. Redžić, "The case of the Doppler effect for photons revisited," Eur. J. Phys. **34**, 1355-1366(2013).

20. E. Schrödinger, "Dopplerprinzip und Bohrsche Frequenzbedingung," Phys. Zeits. **23**, 301-303(1922).

21. E. G. Dwight, *American Institute of Physics Handbook* (2nd Ed.), (New York, McGraw-Hill,1963)

22. I. R. Senitzky, "Sidebandes in strong-field resonance fluorescence," Phys. Rev. Lett. **40**, 1334-1336(1978).

23. G. S. Agarwal, A. C. Brown, L. M. Narducci, G. Vetri, "Collective atomic effects in resonance fluorescence," Phys. Rev. A **15**, 1613-1624(1977).

24. G. S. Agarwal, R. Saxena, L. M. Narducci, D. H. Feng, R. Gilmore, "Analytical solution for the spectrum of resonance fluorescence of a cooperative system of two atoms and the existence of additional side-bands," Phys. Rev. A **21**, 257-259(1980).

25. R. J. Glauber, "The Quantum Theory of Optical Coherence," Phys. Rev. **130**, 2529-2539(1963).

26. F. T. Arecchi, E. Gatti, A. Sona, "Time distribution of photons from coherent and Gaussian sources," Phys. Lett. **20**, 27-29(1966).

27. M. Titze, H. B. Li, "Interpretation of optical three-dimensional coherent

spectroscopy," Phys. Rev. A **96**, 032508(2017).

28.S. G. Yu, M. Titze, Y. F. Zhu, X. J. Liu, and H. B. Li, "Observation of scalable and deterministic multi-atom Dicke states in an atomic vapor," Opt. Lett. **44**, 2795-2798(2019).